\documentclass{IEEEcsmag}

\usepackage[colorlinks,urlcolor=blue,linkcolor=blue,citecolor=blue]{hyperref}

\usepackage{listings}
\usepackage{xcolor}

\newcommand{\colortext}[1]{{\color{black}#1}}
\newcommand{\recolortext}[1]{{\color{black}#1}}

\usepackage{caption}
\DeclareCaptionFormat{listing}{#1#2#3}
\colorlet{punct}{red!60!black}
\definecolor{background}{HTML}{EEEEEE}
\definecolor{delim}{RGB}{20,105,176}
\colorlet{numb}{magenta!60!black}

\definecolor{eclipseStrings}{RGB}{42,0.0,255}
\definecolor{eclipseKeywords}{RGB}{127,0,85}
\colorlet{numb}{magenta!60!black}

\lstdefinelanguage{json}{
	basicstyle=\normalfont\ttfamily\scriptsize,
	numbers=left,
	numberstyle=\scriptsize,
	stepnumber=1,
	numbersep=1pt,
	showstringspaces=false,
	breaklines=true,
	frame=lines,
	backgroundcolor=\color{background},
	tabsize=2,
	commentstyle=\color{eclipseStrings}, 
	stringstyle=\color{eclipseKeywords}, 
	string=[s]{"}{"},
	comment=[l]{:\ "},
	morecomment=[l]{:"},
	literate=
	*{0}{{{\color{numb}0}}}{1}
	{1}{{{\color{numb}1}}}{1}
	{2}{{{\color{numb}2}}}{1}
	{3}{{{\color{numb}3}}}{1}
	{4}{{{\color{numb}4}}}{1}
	{5}{{{\color{numb}5}}}{1}
	{6}{{{\color{numb}6}}}{1}
	{7}{{{\color{numb}7}}}{1}
	{8}{{{\color{numb}8}}}{1}
	{9}{{{\color{numb}9}}}{1}
}

\usepackage{upmath}

\jvol{XX}
\jnum{XX}
\paper{8}
\jmonth{May/June}
\jname{IT Professional}
\pubyear{2019}

\begin{document}

\sptitle{Department: Head}
\editor{Editor: Name, xxxx@email}

\title{Crowd Sensing and Living Lab Outdoor Experimentation Made Easy}

\author{Evangelos Pournaras}
\affil{School of Computing, University of Leeds, UK}

\author{Atif Nabi Ghulam and Renato Kunz}
\affil{Computational Social Science, ETH Zurich, Switzerland}

\author{Regula H{\"a}nggli}
\affil{Department of Communication \& Media Research (DCM), University of Fribourg, Switzerland}

\markboth{E. Pournaras, A. N. Ghulam, R. Kunz, R. H{\"a}nggli}{Crowd Sensing and Living Lab Outdoor Experimentation Made Easy}

\begin{abstract}
\colortext{\recolortext{Living lab outdoor} experimentation using pervasive computing provides new opportunities: higher realism, external validity and \recolortext{socio-spatio-temporal observations in large scale}. However, experimentation `in the wild' is complex and costly. Noise, biases, \recolortext{privacy concerns, compliance with standards of ethical review boards}, remote moderation, control of experimental conditions and equipment perplex the collection of high-quality data for causal inference. This article introduces Smart Agora, a novel open-source software platform for rigorous systematic outdoor experimentation. Without writing a single line of code, highly complex experimental scenarios are visually designed and automatically deployed to smart phones. Novel geolocated survey and sensor data are collected subject of participants verifying desired experimental conditions, for instance, their \recolortext{localization} at certain urban spots. This new approach drastically improves the quality and purposefulness of crowd sensing, tailored to conditions that confirm/reject hypotheses. The features that support this innovative functionality and the broad spectrum of its applicability are demonstrated. }
\end{abstract}

\maketitle

\chapterinitial{Is the future} of scientific human experimentation outdoors? Pervasive computing shapes this future on the basis of a large market for smart phone hardware and software. Ubiquitous devices turn into living laboratory instruments awaiting to directly collect real-world observations at massive scale to tackle some of the most challenging research questions and \colortext{hypotheses~\cite{Hossain2019,Oliver2020,Nanni2021}. In this future of Smart City computing applications such as transportation, mobility and sharing \recolortext{economies}, new epistemological and methodological challenges~\cite{Salganik2019} arise for social and environmental science, economics or psychology.} For instance, how to practice causal inference, how to mitigate trade-offs of realism vs. environment controllability, how to handle biases and counterfactuals~\cite{Arechar2018}? 

The COVID-19 pandemic makes these questions timely and pressing given the safety measures against infection risk in indoor laboratory environments. Are existing online approaches~\cite{Arechar2018} cost-effective for outdoor ubiquitous experimentation? 

\recolortext{
\subsection{Experimentation `in the wild'? Not easy}\label{subsec:related-work}
}

Several existing efforts focus on online web experimentation. For instance, \colortext{nodeGame is a Javascript-based tool for online real-time synchronous decision games and experiments ~\cite{Balietti2017}}. Similarly, z-Tree is a client-server software for ready-made economic experiments~\cite{Fischbacher2007} and oTree is a Python framework that supports multiplayer or single-player strategy games, dynamic questionnaires and assessments/tests~\cite{Chen2016}. Experimentation based on pervasive computing includes AWARE, an open-source mobile instrumentation toolkit that focuses on the resource constraints of mobile devices~\cite{Ferreira2015}. StimuliApp is designed to create psychophysical tests with precise timing on iOS and iPad devices~\cite{Marin2020}. A simple system of menus can support the design and running \colortext{of} tests that assess perceptual capabilities. \recolortext{Another way to collect geolocated crowd sensing data is the so called `digital public notice areas', however, their use requires specialized software/hardware and they mainly support commercial applications~\cite{Alt2011}.} The AIRBOT mobile information assistant is proposed for complex tasks in public space. It has been used to support air passengers in successfully boarding on a plane~\cite{Kilian2019}. 


Several limitations are identified in these existing approaches: (i) The design of an experiment can be complex and usually requires technical/programming skills to achieve rigor or implement a more systematic collection of high-quality data. (ii) A seamless integration to real-world outdoor activities is usually limited. (iii) No general platform approach to outdoor experimentation - pervasive system tools are limited to specific experimental and application scenarios. (iv) Limited automation support for both designers of experiments and participants. (v) Limited open-source and well-documented approaches for outdoor human experimentation. 

\recolortext{
\subsection{How to make it easier? The Smart Agora}\label{subsec:make-it-easy}
}

This article addresses these challenges by introducing the solution of Smart Agora (\url{https://smart-agora.org}). \colortext{Designers visually create complex crowd sensing experimental scenarios on a web browser and deploy them to run without moderation to the smart phones of participants. The collected data are of high quality and socio-spatio-temporal in nature, purposeful to minimize privacy cost, noise and biases. The living lab experimental processes of Smart Agora are tailored by design to satisfy desired experimental conditions, for instance, geolocated survey responses verified by location proofs of participants at certain urban spots\recolortext{~\cite{Pournaras2020}, e.g. on the blockchain}. The broad applicability of Smart Agora is found at several Smart City domains such as cycling risk, urban infrastructures, decisions for participatory budgeting projects and other~\cite{Pournaras2020,Castells2019,Hanggli2021}. }

The contributions of this article are summarized as follows:

\begin{itemize}
	\item A flexible and modular modeling architecture for crowd sensing and outdoor living lab experimentation that supports a large spectrum of complex experimental scenarios. 
	\item The open-source platform of Smart Agora that realizes the modeling architecture by implementing the Smart Agora Dashboard for designers of experiments and the Smart Agora App for recruited participants.
	\item \colortext{A review of Smart Agora use cases to demonstrate the broad spectrum of its applicability.}
	\item A software artifact demonstrator~\cite{SmartAgoraArtifact2021} of Smart Agora running on a virtual machine for reproducability, engagement and assessment of the approach by the broader community. 
	\item Systematic documentation of Smart Agora for users and developers. 
\end{itemize}

The `Software Accessibility' is outlined at the end of this article.

\section{The Smart Agora Living Lab}\label{sec:smart-agora} 

The Smart Agora living lab is an open-source crowd sensing framework for visually designing and carrying out complex outdoor socio-spatio-temporal experiments. It consists of the \emph{Smart Agora Dashboard} and the \emph{Smart Agora App}. The Smart Agora platform is documented and can be easily tested in a provided virtual machine~\cite{SmartAgoraArtifact2021}. \colortext{All these are listed in `Software Accessibility', at the end of this article.}

\recolortext{
\subsection{Experimental design: Smart Agora Dashboard}\label{subsec:dashboard}

The Smart Agora Dashboard} is a web platform designed to simplify the creation of complex outdoor experimental scenarios using pervasive devices such as smart phones. The designer of Smart Agora experiments creates geographical \emph{points of interest} where participants share experimental data, e.g. geolocated survey and sensor data. \recolortext{This dashboard} provides the option to reuse and modularize several complex data collection processes as well as deploy them at different time points with different participants. The design of experimental scenarios is visual and interactive without the need to write a single line of code. However, each experimental scenario can be fully encoded and parsed within JSON files that \colortext{allow} automated generation of complex experiments. For instance, consider experiments \colortext{that require the collection of data} at every bus or tram station in a city. The points of interests can be automatically loaded by including the location of the stations in the JSON files, along with the necessary data collection configurations. The following JSON 1 code listing illustrates an example of how data collection is encoded. 

\begin{lstlisting}[caption=Example of a data collection asset. It contains configurations for a point of interest coming with a survey question with two options to choose from as well as two sensor data collections (gyroscope and GPS location). , language=json,firstnumber=1]
	{
		"Id": "AXeG00HIQMa8aD8nfimV",
		"Name": "Simple_09022021_134527",
		"Url": "http://smart-agora.org",
		"Metadata": {
			"record": {
				"StartAndDestinationModel": {
					"StartLatitude": null,
					"StartLongitude": null,
					"DestinationLatitude": null,
					"DestinationLongitude": null,
					"Mode": "Simple",
					"DefaultCredit": "3"
				},
				"SampleDataModel": [{
					"id": 1,
					"Question": "How dangerous for bikers was the last section?	",
					"Type": "radio",
					"Latitude": "47.3715915",
					"Longitude": "8.538603799999999",
					"Sensor": [{
						"id": 1,
						"Name": "Gyroscope"
					},{
						"id": 2,
						"Name": "Location"
					}],
					"Time": "3",
					"Frequency": "Medium",
					"Sequence": "Disable",
					"Visibility": "true",
					"Mandatory": "true",
					"Option": [{
						"id": 1,
						"Name": "Safe",
						"NextQuestion": null,
						"Credits": ""
					},{
						"id": 2,
						"Name": "Dangerous",
						"NextQuestion": null,
						"Credits": ""
					},],
					"Combination": null,
					"Vicinity": "25"
				}]
			}
	}}
\end{lstlisting}

\recolortext{
	\subsection{Geolocated data collection: Smart Agora App}\label{subsec:app}
}

\begin{figure}[!htb]
	\centering
	\includegraphics[width=0.49\columnwidth]{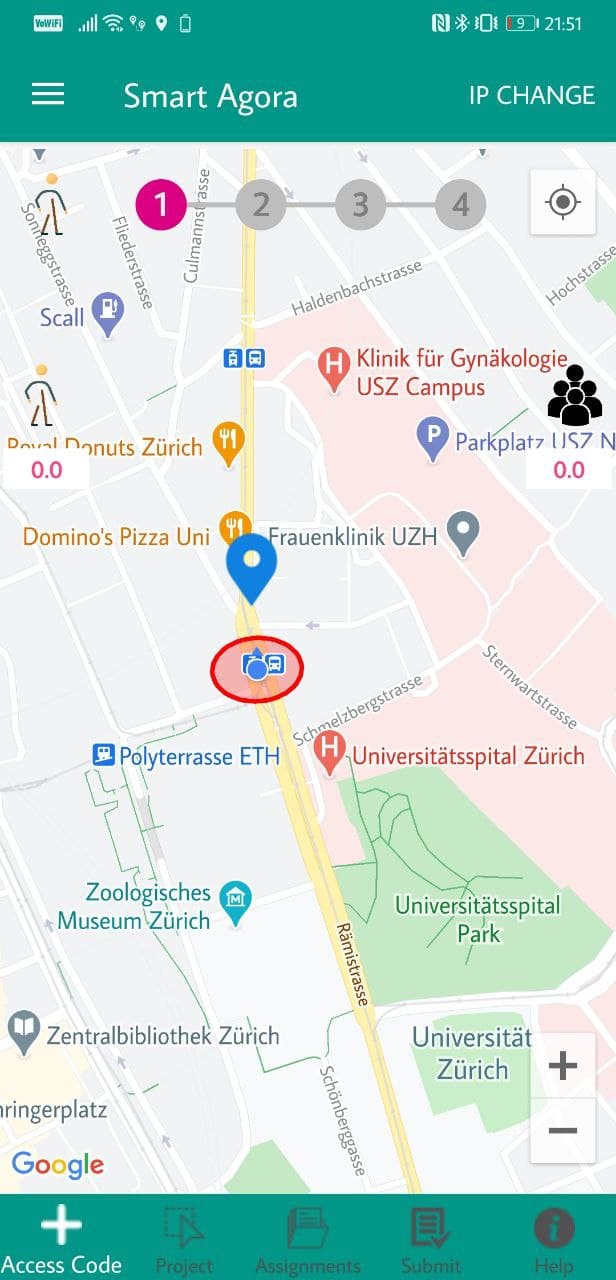}
	\includegraphics[width=0.49\columnwidth]{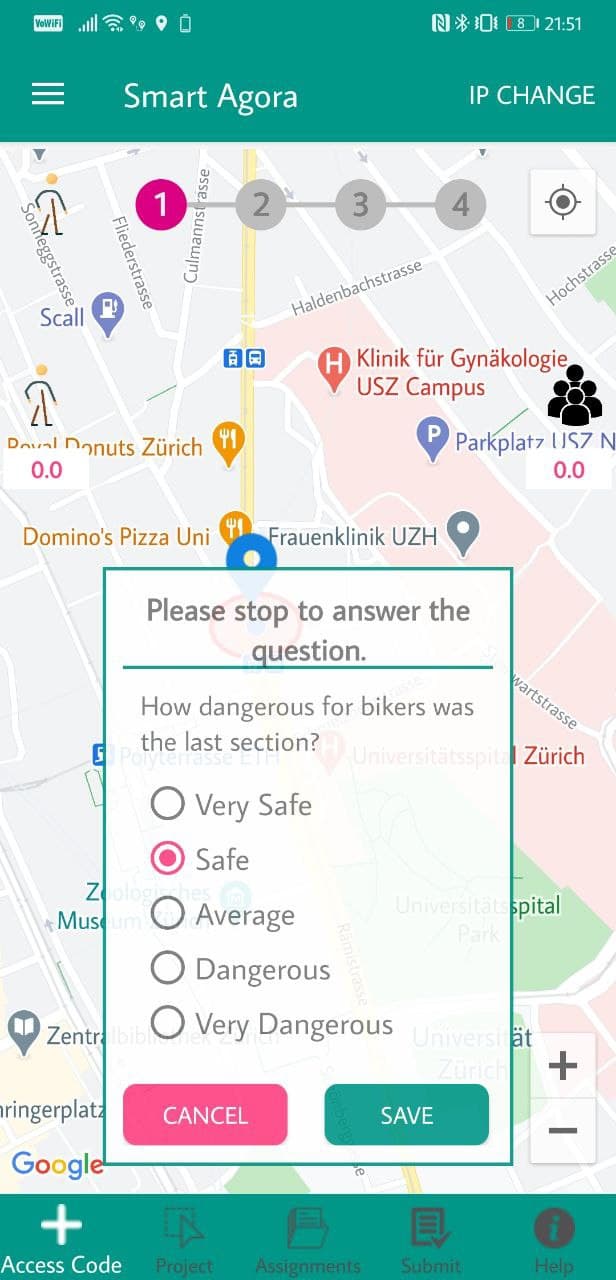}
	\caption{The augmented map of the Smart Agora App.}\label{fig:smart-agora-app}
\end{figure}

The Smart Agora app is an Android app that allows participants to subscribe and participate to experiments. These experiments are loaded in an augmented map that contains the points of interest that participants visit to answer questions on spot, while sharing smart phone sensor data (see Figure~\ref{fig:smart-agora-app}). Visiting the points of interests is verified when participants are in close proximity with a point of interest measured by a radar-like radius of a localization cycle. A similar method with an ellipse is also supported, where the shortest axis of the ellipse points to the direction of the next point of interest~\cite{Griego2017}. This elliptical method of localization could be suitable in urban environments with narrow streets. Smart Agora localization currently operates based on GPS technology. However, approaches based on distributed ledgers operating with LoRaWAN and adhoc networks can be more secure and reliable alternatives~\cite{Pournaras2020}. 

\recolortext{
	\subsection{Innovations and special features}\label{subsec:innovations}
}

Smart Agora supports by design innovative experiments that improve data collection quality. Compared to continuous privacy-intrusive sensing approaches~\cite{Rabbi2020}, Smart Agora collects \colortext{minimal data, tailored} to specific context and purpose and as such it is easier for the \recolortext{designers} of the experiments to meet minimal data collection requirements of ethical review boards. This is made possible by making participants' responses subject of \emph{proving witnessed presence}~\cite{Pournaras2020}. Answering survey questions and sharing sensor data is performed after verification of the location or another criterion related to the situational awareness of participants using QR codes, challenge questions, puzzles and CAPTCHA-like tests. This is useful in experiments with outdoor group interactions in which group members require to prove their identity and membership to each other group member. Similarly, \recolortext{the observations of experiments involving physical objects}, e.g. the effect of a political poster, the response of a traffic light for pedestrians or arrival delays in a bus \recolortext{station}, could be verified with a QR code on the physical object. These verification processes result in automating the control of experimental conditions that is usually performed manually by lab moderators in indoor lab environments. 


\recolortext{
	\subsection{Fields of application}\label{subsec:applicability}
}

\colortext{Proving witnessed presence has a significant impact on improving the quality of data collection. Data sharing is designed as responsible testimonies and citizens' interventions~\cite{Pournaras2020}. This makes Smart Agora highly \recolortext{versatile and} applicable in different living lab manifestations:} (i) \emph{Science} - scientists design experiments and participants are recruited to participate. \recolortext{Cycling risk} assessments is an example here~\cite{Castells2019,Pournaras2020}. (ii) \emph{Policy-making} - policy makers design participatory decision-making and deliberation processes, while citizens participate directly in crowd-sourced decisions and \recolortext{bottom-up} policy formation. \recolortext{The participatory budgeting field test} in the city of Aarau is an example here~\cite{Hanggli2021}. (iii) \emph{Education} - teachers and lecturers design highly engaging outdoor learning activities and students learn out of the class. The 2018 MSc course ``Data Science in Techno-socio-economic Systems" at ETH Zurich is an example here.



\section{Smart Agora Crowd Sensing Modeling}\label{sec:modeling}

\begin{figure*}[!htb]
	\centering
	\includegraphics[width=1.0\textwidth]{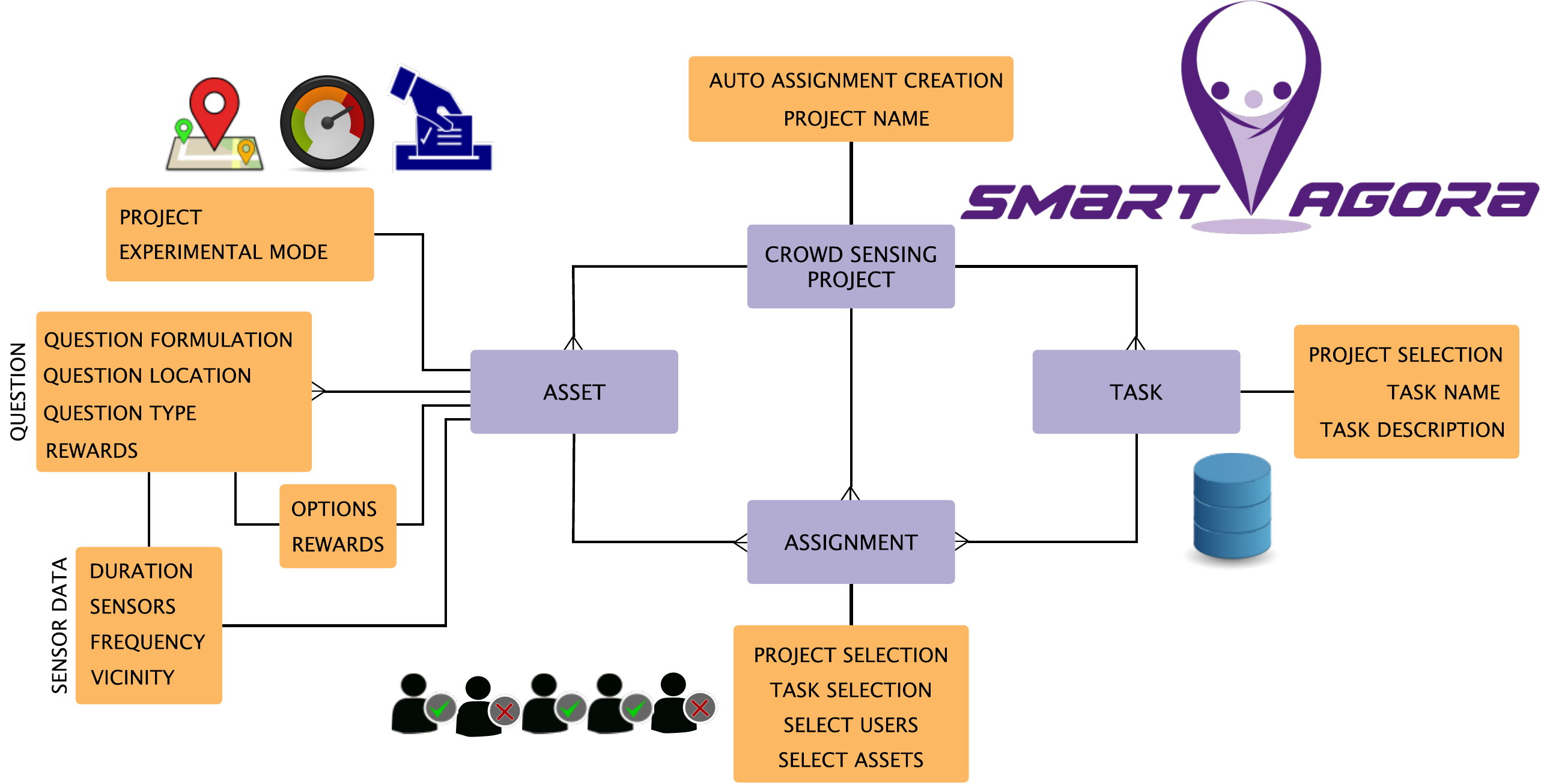}
	\caption{The Smart Agora modeling architecture.}\label{fig:architecture}
\end{figure*}

Figure~\ref{fig:architecture} introduces the modeling architecture of Smart Agora. It consists of four core elements: (i) \emph{project}, (ii) \emph{asset}, (iii) \emph{task} and (iv) \emph{assignment}. A \recolortext{crowd sensing} project can contain several data collection assets that encode experimental scenarios. Each asset can be used in several unique data collection processes by creating a task and linking this task with the asset via an assignment to participants. A task results in data collected by the assigned participants according to an asset. All experimental processes can \colortext{be} designed visually without writing a single line of code. The Smart Agora architecture is implemented by \recolortext{the} \texttt{hive} platform built in Go and Elasticsearch. 

\begin{figure*}[!htb]
	\centering
	\includegraphics[width=0.75\textwidth]{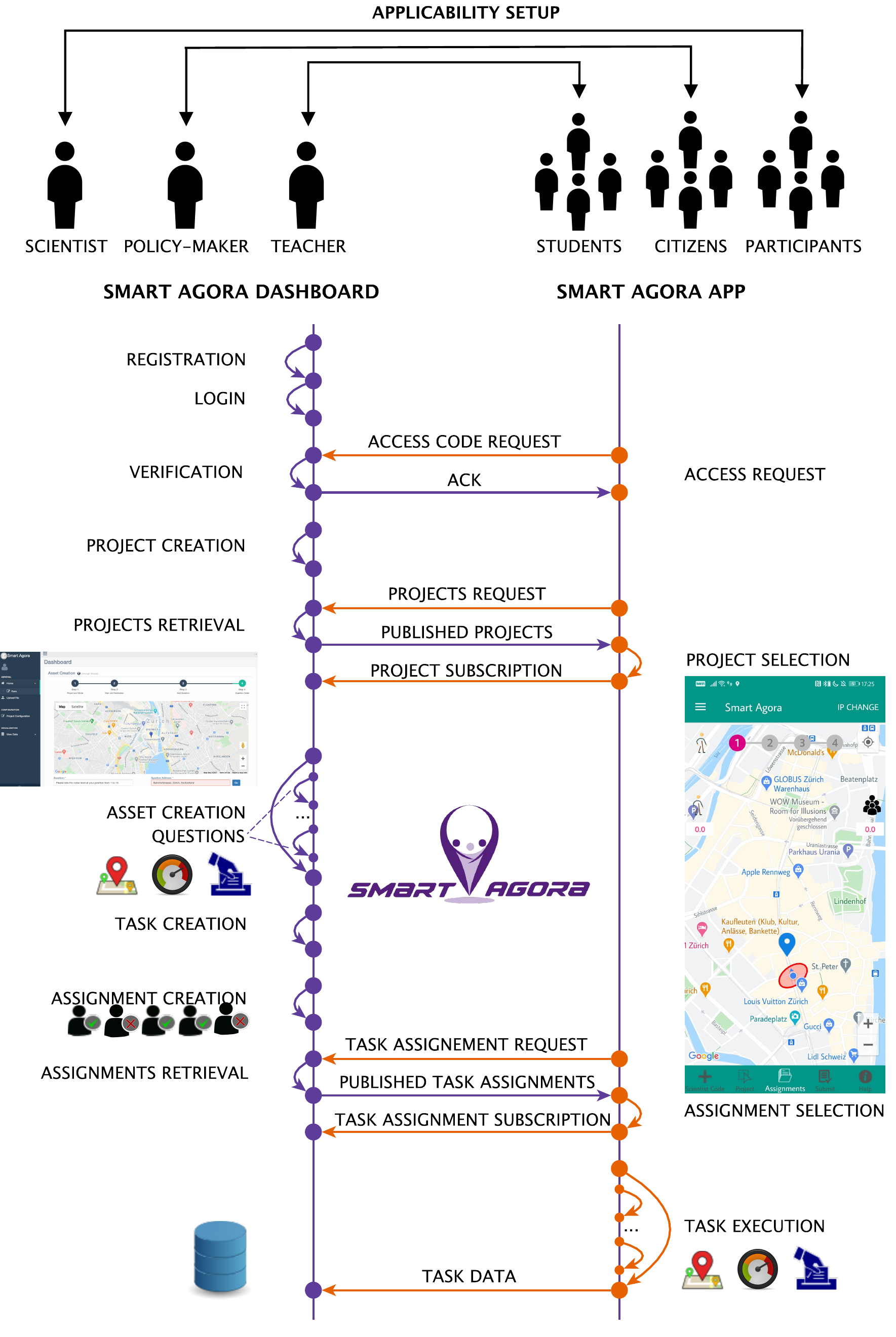}
	\caption{\colortext{The lifecycle of a Smart Agora experiment.}}\label{fig:lifecycle}
\end{figure*}

Figure~\ref{fig:lifecycle} illustrates the lifecycle of a Smart Agora experiment from design \recolortext{to} deployment and completion of the data collection. The designer of the experiment creates an account to the Smart Agora Dashboard and logins to create a crowd sensing project. \colortext{The participants of an experiment download the Smart Agora App and are granted access to the subscription of projects using an access code provided to them by the moderators of the experiment as part of the recruitment process. }

In the meantime, the designer of the experiment creates data collection assets that contain the points of interests with the survey questions and the \colortext{desired mobile sensor data for collection}. These assets are associated with tasks via the creation of assignments that control which participants can join the experiment. The participants can subscribe to the task assignments. \colortext{This loads} the points of interests and survey questions to their smart phone map, while it configures the sensor data collection, e.g. frequency. Upon completion of the task, the data are transferred to the Smart Agora database for storage and analysis. Smart Agora also provides the option to forward collected data to DIAS (\url{http://dias-net.org}), the \emph{Dynamic Intelligent Aggregation Service}~\cite{Pournaras2017} for decentralized real-time data analytics~\cite{Pournaras2020}. 

The rest of this section illustrates the features of Smart Agora and how they support different outdoor experimental scenarios. 

\subsection{Crowd sensing project}\label{subsec:project}

A Smart Agora crowd sensing project is the top of the modeling hierarchy and facilitates a group of experimental processes and collected data. It is initially created by the designer of the experiment and contains multiple \recolortext{crowd sensing} assets, tasks and assignments. By default, an auto-assignment option connects a created asset with the most recent task and is made available immediately to all available participants in the pool. A crowd sensing project can be used to group together a series of several experiments and collected data that serve a research project. In this way, the whole experimental design of a research project can be built incrementally and managed more efficiently. 

\subsection{Crowd sensing asset}\label{subsec:asset}

The crowd sensing asset provides toolkits to design ubiquitous experiments and the collection of socio-spatio-temporal data. An asset is associated with a project. An experiment can be conducted in \colortext{three navigational modalities}:

\begin{itemize}
	\item \textbf{Simple \colortext{modality}}: Navigation to different points of interests occurs in any order. This experimental \recolortext{modality} is provided for experiments with a large number of points of interests (e.g. all bus stations of a city), probably expanding to a large geographic area of several possible participants. Data collection is usually long-term and \colortext{large in scale, making it suitable to compare experimental observations over different geographical areas and over time, i.e. evolving human behavior and environment. }
	\item \textbf{Sequential \colortext{modality}}: Navigation to different points of interests occurs in a predetermined order introducing in this way a well structured experimental process that has to be followed by participants (e.g. following given vehicle routing trips). \recolortext{The information into which a participant is exposed at a certain point of interest determines the behavior and survey responses at the next points of interests}. This allows to observe and study systematically a socio-spatio-temporal phenomenon (e.g. mobility pattern) by tracing at check points (the points of interests) how it incrementally evolves. \colortext{Compared to the simple navigation modality, this one is suitable for more controlled experimental scenarios with a lower number of points of interests and participants}, with the geographical scale depending on the transport mode involved, e.g. walking pedestrians vs. cyclists, vs. vehicle drivers. 
	\item \textbf{Dynamic \colortext{modality}}: Navigation to different points of interests is determined by survey responses \colortext{or by detected events in the collected sensor data}. The next point of interest to visit is determined by the answer to a question at a previous \colortext{point of interest}. This feature allows the design of `what-if' and gamification scenarios \colortext{(`serious games')}  in which participants are involved in an outdoor living lab problem solving. \colortext{An experiment conducted under the dynamic modality can be used to vary on-the-fly experimental conditions of participants based on their demographics or environmental variables they experience and detected within the sensor data.}
\end{itemize}

Moreover, an asset consists of multiple geolocated survey questions. Each question consists of a location on a map (GPS coordinates), the question formulation, the question type (radio, checkbox, likert scale and text box) and an optional amount of rewards that will be provided to the participants for answering this question. These rewards can represent the recruitment compensation and incentives to preserve engagement or scale up data collection, while minimizing attrition effects~\cite{Arechar2018}. Rewards can be monetary, discount coupons or capture value measured with a designed cryptocurrency (token). Smart Agora can couple with other systems to automatically manage transactions and the actual secure value exchange, for instance digital wallets that verify transactions on distributed ledgers (blockchain). Depending the question type, the designer may introduce the options to choose from. Additional rewards may be provided for each selected option to allow experimental economics research. 

Each point of interest can be further augmented with the collection of smart phone sensor data including sensors such as:  light, gyroscope, proximity, accelerometer, GPS location and noise. Sensor data are collected in a determined vicinity around the point of interest during a determined duration and at \colortext{a low (every 2 sec), medium (every 250 ms) or high (every 200 ms) frequency}. \colortext{A large localization radius can provide unrestricted collection of sensor data in experimentation scenarios requiring a continuous and passive crowd sensing data.} \colortext{Sensor data and sensor fusion can be used for human activity recognition (traveling vs. walking vs. cycling) as well as social inferences (who speaks with whom on spot), including the inferences to prove witnessed presence~\cite{Pournaras2020}.}

\colortext{The creation of an asset is a collection of multiple points of interests configured to collect any combination of survey question \recolortext{data} and/or smart phone sensor data}. They are created interactively via a visual wizard that guides the designer step by step to build a crowd sensing asset. \recolortext{At the end, this wizardl generates files, such as the JSON 1 code,} that represent the experiment and the designer's choices. Several assets can represent different control and treatment groups for comparisons.

\subsection{Crowd sensing task}\label{subsec:task}

A crowd sensing task is determined within a project. It is associated with an asset via an assignment and as such it contains the collected experimental data. Tasks allow a degree of freedom in the data that can be collected by (re)using one or more assets. For instance, given a crowd sensing asset, the same data can be collected at different time points by the same or different groups of participants. And vice versa, the same data can be collected by different groups of participants at the same or different time points.

\subsection{Crowd sensing assignment}\label{subsec:assignment}

An assignment connects an asset and a task to the participants of an experiment. The designer has control of which participants to choose from the ones recruited and available in the pool. 

\colortext{
\section{Smart Agora in Action: Use Cases}\label{sec:use-cases}

Four different use cases are outlined here to demonstrate the applicability spectrum of Smart Agora and its impact on outdoor experimentation. 

\subsection{Crowd sensing of cycling safety }\label{subsec:cycling}

The capability of Smart Agora to facilitate informed decisions by geolocating the decision-making process (proof of witnessed presence) is assessed in a study on cycling safety~\cite{Pournaras2020}. The perception of bike riders about the cycling safety of different urban spots in Zurich is compared to an empirical safety model~\cite{Castells2019} built using official data of the Federal Roads Office collected from Swiss GeoAdmin. If the two safety estimations match, this is indication that witnessed presence in participatory crowd sensing can indeed provide information quality comparable to the official but costly data collection methods.

The empirical risk estimation of a route is derived by a continuous spatial accident risk model of Zurich~\cite{Castells2019}. This is the baseline that is compared to the perceived cycling risk estimated via 11 Smart Agora test users. A crowd sensing asset is designed using the sequential navigation modality with four points of interest. \recolortext{The cycling risk of each road section that connects two consecutive points of interests is assessed at the latter point}. A likert scale question pops up in the Smart Agora app to \recolortext{evaluate} cycling risk at a linear scale between 1. \emph{very safe} to 5. \emph{very dangerous}. All test users cycled at the same day and time to minimize biases originated from weather, light condition and the condition of different bikes. 

A high matching is found between the empirical evidence from the cycling accident data and the cycling risk that citizens witness~\cite{Pournaras2020}. This motivates an alternative more legitimate policy-making approach that relies on high-quality crowd sensing data collected by verifying witnessed presence.

\subsection{Geolocated voting for participatory budgeting}\label{subsec:participatory-budgeting}

\recolortext{Along with collecting high-quality policy-making data, Smart Agora can also empower novel participatory processes that have earlier been too costly or complex for digital democracy.} For instance, a participatory budgeting field test is planned to run in the city of Aarau in Switzerland, where citizens will decide themselves for a Smart City project to fund using a municipal budget of 10K CHF to improve their life~\cite{Hanggli2021}. Examples of such projects include transport decarbonization initiatives (e.g. bike sharing stations), digital cultural heritage, smart waste bins and other. 

The use of Smart Agora for outdoor experimentation opens up several new research avenues that will be explored in this project. For instance, to what extent proving witnessed presence results in more legitimate voting outcomes? Letting citizens of a local district community to witness urban spots that will be transformed by Smart City projects has an added value at different levels: Citizens can physically come closer to each other to deliberate about the impact of the project on the community and urban surrounding. Citizens can improve their awareness by collecting evidence to support a more informed and responsible decision. This process reunites the digital and physical decision space of citizens, inspiring the revive of a new cyber-physical `ancient Agora of Athens'.  Ultimately, this process shall result in a voting outcome accepted by the broader community and even by citizens that voted against the elected project~\cite{Hanggli2021}. 

Moreover, voting rules, e.g. majority voting vs. multi-option preferential voting, can influence voting outcomes as well as the capacity of local communities to reach legitimate decisions that promote consensus. Smart Agora can implement different voting rules applied to different treatment groups during the field test to make novel comparisons in the context of outdoor geolocated decision-making. 

\subsection{Decentralized sensing of transport sustainability }\label{subsec:DIAS}

Beyond the persistence of collected data in the Smart Agora database for offline analysis, the integration of DIAS to Smart Agora provides \recolortext{capabilities for decentralized real-time data analytics}~\cite{Pournaras2020}. In particular, collective measurements (summation, average, max, min, count, etc.) can be made and updated in real-time based on the most recent shared data of participants localized at points of interest. When participants depart from a point of interest, automated rollback operations are performed to preserve accurate estimations using exclusively localized participants. These novel calculations are performed in an interactive and cooperative fashion by participants' devices and without the involvement of any centralized third party. DIAS implements the distributed communication protocol and a memory system for efficient localized data management, see earlier work for technical details~\cite{Pournaras2017,Pournaras2020}. 

To demonstrate this progressive capability, a complex field test was conducted in Zurich with 6 test users to assess the sustainability of transport usage ~\cite{Pournaras2020}. The participants choose the preferred transport mean to reach two points of interest using a simplified linear model of sustainability: 0. \emph{Car}, 1. \emph{Bus}, 2. \emph{Train}, 3. \emph{Tram}, 4. \emph{Bike}, 5. \emph{Walking}. \recolortext{The mean sustainability is updated in real-time as participants arrive and depart from the points of interest, where they choose their preferred transport mean, see the video at `Software Accessibility'.  The observed estimations of transport sustainability approximate well the actual sustainability values. This confirms the efficiency of Smart Agora to carry out decentralized real-time collective measurements.}


\subsection{COVID-19 and field experimentation}\label{subsec:COVID-19}

The COVID-19 pandemic creates several challenges for indoor experimentation. Participants and moderators are exposed to significant infection risks. Several university labs have been closed during lockdowns, while setting up a safe lab can be costly, complex and time-consuming, requiring institutional approvals along with new frameworks of safety and operation (ventilation, disinfection, COVID-19 testing, recruitment, ethical approvals, etc.).

 In contrast, outdoor experimentation with Smart Agora reduces infection risk for participants as experimental tasks are deployed automatically and remotely. Smart Agora can also provide a higher degree of freedom to scale up recruitment and ubiquitous data collection, lower moderation cost and higher realism. Moving experimentation outdoors provides new opportunities to limit sampling biases of indoor lab pools of participants, often overpopulated by young individuals with a student profile. Participants from older age groups can be engaged given the priority they have to vaccination as well as the opportunity to eliminate isolation of elderly during lockdown and to contribute to public good. 

Recruitment for experimentation can be heavily affected due to COVID-19 governmental constraints on mobility. These can restrict people to move only in close proximity for essential purposes such as going to work, shopping or exercising. Other activities may be prohibited and even fined. Smart Agora provides new opportunities to seamlessly integrate experimental processes within permitted mobility ranges. For instance, instead of an indoor driving simulator, driving behavior can be studied during pandemics when citizens drive to work or for shopping by implementing a travel diary using Smart Agora. Similarly, Smart Agora can also serve COVID-19 research~\cite{Nanni2021,Oliver2020}, for instance understanding infection risk and epidemic spread by crowd sensing collective human behavior and actions. 

Incentives to participate in experimentation may deviate from the norm during pandemics. Lab recruitment can be a flexible income support given the loss of jobs. Non-monetary incentives have a potential to strengthen the sense of community belonging using participatory and gamification elements~\cite{Seaborn2015} such as the ones of Smart Agora. These incentives may capture new ways to trigger intrinsic human values during lockdowns, for instance, higher interest in contributing to science, i.e. citizen science, outdoor exercising or well being. 
}

\section{Broader and Future Perspective}\label{sec:discussion}




\colortext{As Smart Cities co-evolve to complex ecosystems of compelling socio-technical complexity, \recolortext{living lab outdoor experimentation} will remain a timely and significant challenge to tackle. Collecting crowd sensing data is not only the end goal but also the mean to reason about desired experimental conditions to carry out rigorous studies involving causal inferences. Verifying such conditions in an automated way opens up new avenues to carry out large-scale outdoor experimentation with lower biases, noise and moderation level. Smart Agora is an endeavor along these lines, made to support both experiment designers and participants in different manifestations of science, policy-making or education. }

Several challenges though remain such as how to incentivize participation and minimize dropouts as well as identify and minimize unknown biases that will emerge in complex urban environments. Outdoor experimental participation requires new intrinsic values for incentivizing citizens' contributions to science. Distributed ledgers and cryptoeconomic principles such as token curated registries are promising means to capture and securely exchange such values. \colortext{This is also expected to scale up the number of participants in future use cases. }

\colortext{Despite the minimal and targeted data collection approach of Smart Agora, solutions for privacy protection and data sharing control will be implemented in future, for instance, differential privacy and homomorphic encryption to privately aggregate sensitive citizens' data. Participants may require self-determined access control rights to sensor data to deliver certain crowd sensing tasks~\cite{Pournaras2017}.} \recolortext{Hence, data collectors need to match with data suppliers based on shared values that go beyond monetary ones and resolve social dilemmas related to privacy, ethics, quality of life, comfort and other. For instance, citizens' assemblies may play a role of ethical review boards determining experimental participation and as such paving the way towards new standards and norms on how to conduct experimental research. This will allow citizens to `shape' and ethically align experiments to their own common values and in a bottom-up way, pushing towards a more socially responsible science.}

\section{Software Accessibility}\label{sec:software}

The Smart Agora platfrom \colortext{and supporting material} are open-source and accessible via the links below: 

{\footnotesize
\begin{itemize}
\item Smart Agora Dashboard:\\\url{https://github.com/epournaras/SmartAgoraDashboard}
\item Smart Agora App:\\\url{https://github.com/epournaras/SmartAgoraApp}
\item Smart Agora Documentation:\\\url{https://epournaras.github.io/SmartAgoraDocumentation}
\item Tutorial of the Smart Agora demonstrator~\cite{SmartAgoraArtifact2021}:\\\url{https://www.youtube.com/watch?v=6KYjd4AaNkE}
\item \colortext{Smart Agora - DIAS Demonstrator:\\\url{https://www.youtube.com/watch?v=ULSMypnvL9w}}
\item The \texttt{hive} platform \colortext{on which Smart Agora relies}:\\\url{https://github.com/nytlabs/hive}
\end{itemize}
}

\section{ACKNOWLEDGMENT}

This research is funded by the SNF NRP77 `Digital Transformation'
project “Digital Democracy: Innovations in Decision-making Processes”,  \#407740 187249: \url{https://www.nfp77.ch/en/portfolio/trust-and-legitimation-in-the-digital-democracy/}. The authors would also like to thank Prof. Dirk Helbing and the Chair of Computational Social Science for supporting this research. Edward Gaere supported the integration of Smart Agora with DIAS~\cite{Pournaras2017}, while Dr. Alexey Gokhberg provided the Hive interfaces to Smart Agora. The authors would also like to thank the 2018 students of the course ``\emph{Data Science in Techno-socio-economic Systems}'' at ETH Zurich who used Smart Agora and provided invaluable feedback. Many thanks go to other Empower Polis team members as well as the Institute of Science Technology and Policy (ISTP) of ETH Zurich for running the ETH Policy Challenge and providing a venue to cultivate ideas for outdoor living lab experimentation.  

\bibliography{Smart-Agora} 
\bibliographystyle{IEEEtran}

\begin{IEEEbiography}{Dr. E.Pournaras}{\,} is an Associate Professor at School of Computing, University of Leeds, UK. He is also a research associate at UCL Center of Blockchain Technologies and an Alan Turing Fellow. He has more than 5 years of research experience at ETH Zurich and EPFL in Switzerland including industry experience at IBM T.J. Watson Research Center in the USA. Evangelos received his PhD from Delft University of Technology in 2013. Since 2007 he holds a MSc with distinction in Internet Computing from University of Surrey in the UK and he received his BSc on Technology Education and Digital Systems from University of Piraeus in Greece in 2006. Evangelos has won the Augmented Democracy Prize, the 1st prize at ETH Policy Challenge as well as 5 paper awards and honors. He has published more than 75 peer-reviewed papers. He is the founder of the EPOS, DIAS, SFINA and Smart Agora projects and has worked in several EU/national projects. Evangelos' research interest is on distributed and intelligent social computing systems with expertise in the domains of Smart Cities and Smart Grids.
\end{IEEEbiography}

\begin{IEEEbiography}{R. Kunz}{\,} is currently a Software Developer at Supercomputingsystems in Zurich, Switzerland. He holds a BSc in Computer Science from ETH Zurich since 2021 and he worked at the Chair of Computational Social Science as a Research Assistant during which he studied and developed digital voting systems. He is specialized in mobile and web application development. 
\end{IEEEbiography}

\begin{IEEEbiography}{A.G. Nabi}{\,} is currently a Cloud DevOps Engineer at Swisscom AG in Zurich, Switzerland. Since 2020, he holds a MSc on Informatics from University of Zurich and he has worked as a Research Assistant at the Chair of Computational Social Science at ETH Zurich. His interest lie on distributed artificial intelligence, blockchain/cryptocurrencies, software systems and cloud computing.
\end{IEEEbiography}

\begin{IEEEbiography}{Prof. R. H{\"a}nggli}{\,} is full professor of political communication at the University of Fribourg (Switzerland). From 2011-2013 she was assistant professor at University of Amsterdam. Before, she studied Political Science at the University of Berne, and got a PhD in 2010 from the University of Zurich. Her work deals with (digital) democracy and opinion formation processes, and the interplay between politicians, media, and citizens. She is a Principal Investigator in the NRP 77 “Digital Transformation” with a project about digital democracy. 
\end{IEEEbiography}

\end{document}